# The W. M. Keck Observatory infrared vortex coronagraph and a first image of HIP79124 B

E. Serabyn[1], E. Huby[2], K. Matthews[3], D. Mawet[3], O. Absil[2], B. Femenia[5], P. Wizinowich[5], M. Karlsson[4], M. Bottom[3], R. Campbell[5], B. Carlomagno[2], D. Defrère[2], C. Delacroix[2], P. Forsberg[4], C. Gomez Gonzalez[2], S. Habraken[2], A. Jolivet[2], K. Liewer[1], S. Lilley[5], P. Piron[2], M. Reggiani[2], J. Surdej[2], H. Tran[5], E. Vargas Catalan[4], and O. Wertz[2]

[1]Jet Propulsion Laboratory, California Institute of Technology, 4800 Oak Grove Drive, Pasadena, CA 91109, USA; gene.serabyn@jpl.nasa.gov

[2]Space sciences, Technologies, and Astrophysics (STAR) Institute, Université de Liège, 19c allée du Six Août, 4000 Liège, Belgium

[3]California Institute of Technology, Division of Physics, Mathematics and Astronomy, Pasadena, CA 91125, USA

[4]Department of Engineering Sciences, Ångström Laboratory, Uppsala University, Box 534, 751 21 Uppsala, Sweden

[5]W. M. Keck Observatory, 65-1120 Mamalahoa Hwy, Kamuela, HI, USA

**Abstract**

An optical vortex coronagraph has been implemented within the NIRC2 camera on the Keck II telescope and used to carry out on-sky tests and observations. The development of this new L'-band observational mode is described, and an initial demonstration of the new capability is presented: a resolved image of the low-mass companion to HIP79124, which had previously been detected by means of interferometry. With HIP79124 B at a projected separation of 186.5 mas, both the small inner working angle of the vortex coronagraph and the related imaging improvements were crucial in imaging this close companion directly. Due to higher Strehl ratios and more relaxed contrasts in L' band versus H band, this new coronagraphic capability will enable high-contrast small-angle observations of nearby young exoplanets and disks on a par with those of shorter-wavelength extreme adaptive optics coronagraphs.

# 1. Introduction

Several high-contrast "extreme" adaptive optics (ExAO) imaging systems aimed at the detection of wide-separation Jovian exoplanets have recently come on-line at large ground based telescopes (Beuzit et al. 2008, Macintosh et al. 2014, Jovanovic et al. 2015), all operating at short near-infrared (NIR) wavelengths (1-2.5 μm). On the other hand, longer infrared (IR) wavelengths have significant advantages, including higher Strehl ratios and generally more favorable planet-to-star contrast ratios for young Jovian exoplanets (Burrows et al. 1997, Baraffe et al. 2003, Fortney et al. 2008, Madhusudhan, Burrows & Currie 2011). As a result, even with higher thermal background noise from sky and beam train emission in the L' band (3.8 μm) than at shorter wavelengths, L' planet-mass detection limits for young Jovian exoplanets are in many cases comparable to those attainable with NIR ExAO systems, especially for cooler exoplanets. Indeed, in some cases, observed contrasts are eased by roughly an order of magnitude at L' (Marois et al. 2008, Macintosh et al. 2015). L' high-contrast coronagraphy of self-luminous Jovians, as well as of circumstellar disks, thus holds great promise.

The most important performance parameters describing any coronagraphic system are the attainable contrast ratio, inner working angle (IWA), and sensitivity. The contrast is usually given by the ratio of residual off-axis scattered and diffracted starlight to the peak of the on-axis stellar point spread function (PSF), while the IWA is the smallest angle at which a companion can potentially be detected, usually defined as the angle at which half the planetary signal is transmitted by the coronagraph. The raw on-sky contrast (i.e., the instantaneous image contrast prior to combining or differencing images) attainable at ground-based telescopes is limited by many factors, including diffraction from an on-axis secondary mirror and its support legs, the level of uncorrected (residual) wavefront errors provided by the adaptive optics (AO) system, the pointing accuracy, and the background noise level. Note that in the high thermal-background regime, the best possible performance is set by the background noise. Thus, long-wavelength coronagraphs need only aim at providing sufficient starlight suppression and wavefront correction to enable long integration images to reach the background limit for point-source detection, as in standard photometry.

In order to enable companion searches out to large stellar distances, it is vital to reach the smallest possible angular offsets from host stars. An IWA of $\approx \lambda/D$, where $\lambda$ is the observing wavelength and D the aperture diameter, is 80 mas on a 10-m telescope at L', which can enable exoplanet searches in to ~ 10 AU for nearby star formation regions such as Sco-Cen and Taurus (at 120 - 150 pc), and also observations straddling the ice line (at several AU) for significantly closer stars. A coronagraph with an intrinsically small IWA is critical to enable such observations, thus making phase mask coronagraphs very apropos, as they typically have no intrinsic blockage at small angles. Of these, optical vortex phase masks (Mawet et al. 2005, Foo, Palacios & Swartzlander 2005) have significant advantages (e.g., an IWA of 0.9 lambda/D in the simplest case of a clear circular aperture; the absence of off-axis dead zones; manufacturable masks, etc.), and have proven themselves capable of reaching very small angles both in the lab and on sky (Serabyn et al. 2010, Mawet et al. 2010, 2011, Delacroix et al. 2013, Mawet et al. 2013,

Absil et al. 2013, Serabyn et al. 2013, Defrère et al. 2014, Reggiani et al. 2014, Biller et al. 2014). Moreover, the technology for manufacturing vortex phase masks via form birefringence (sub-wavelength gratings; Mawet et al. 2005) has advanced to the point that high-quality optical vortex phase masks can now be produced with this technique for wavelengths as short as the L' band (Forsberg & Karlsson 2013, Vargas Catalan et al., in press).

Given the combination of higher Strehl ratios (~ 80 – 90% even with first generation AO systems) and more modest intrinsic source contrasts at L', the small IWA of the vortex coronagraph, and the maturity of sub-wavelength-grating technologies, implementation of an L-band vortex coronagraph capability at the W.M. Keck Observatory was seen as both feasible and potentially very scientifically productive, especially as such a system could enable observations of young, hot Jovians to within ~ 80 mas of nearby stars, a regime competitive not only with NIR ExAO systems, but also with aperture masking interferometry to some extent. We have therefore developed an L'-band optical vortex coronagraph on the Keck II telescope, based around the existing adaptive optics (AO) system (Wizinowich et al. 2000) and the NIRC2 camera (http://www2.keck.hawaii.edu/inst/nirc2/). Relying on these existing systems enabled a very rapid and cost-effective implementation. Our approach was thus to quickly develop the new coronagraphic capability as a shared-risk mode, allowing rapid on-sky deployment, leaving a potential upgrade to a facility observational mode to a future phase, if warranted by user interest. This paper briefly describes the new Keck L'-vortex coronagraph, as well as its major development and implementation steps. It also presents an initial demonstration observation of a very close companion enabled by this new observing mode. The main goal of this paper is to provide a rapid overview of this new observational mode. As the development phase is not quite complete, future papers will describe more detailed aspects of the system such as the pointing stabilization approach, the ultimate measured contrast, speckle suppression, and further early science results.

## 2. Performance Expectations

To first order, the integrated leakage, $L$, of starlight after a vortex coronagraph operating behind a centrally obscured telescope aperture can be written as a sum of several dominant contributions, i.e.,

$$L \approx L_s + L_p + L_w, \qquad (1)$$

where $L_s$ is the leakage due to the secondary obscuration, $L_p$ is the leakage due to pointing errors of the star relative to the center of the vortex phase mask, and $L_w$ is the leakage due to wavefront errors other than pointing. The first term is given by

$$L_s \approx \left(\frac{d}{D}\right)^2, \qquad (2)$$

where d is the secondary diameter (Jenkins 2008, Mawet et al. 2013), and the approximation sign is due to the Keck aperture not being circular. Next, for small pointing errors (i.e., < 0.5 λ/D; Huby et al. 2015),

$$L_p \approx \frac{1}{8}\left(\frac{\pi\theta}{\lambda/D}\right)^2, \qquad (3)$$

where $\theta$ is the star's root-mean-square angular offset relative to the vortex center. Finally, for small wavefront errors,

$$L_w = 1 - S \approx \varphi_{rms}^2, \qquad (4)$$

where $S$ is the Strehl ratio (Strehl 1895) and $\varphi_{rms}$ the root-mean-square wavefront phase error (Mahajan 1981, Born & Wolf 2011). For small $\varphi_{rms}$ the stellar PSF can be described as the combination of a fraction $S$ of the light in an Airy pattern, that can be rejected by the vortex, and a fraction 1-$S$ of the light that is scattered out of the PSF core and is largely unaffected by off-center passage through the vortex mask.

Equation 1 allows a comparison of the dominant leakage terms. However, it ignores smaller effects, such as the leakage due to the secondary supports, which could in principle be blocked by a well-matched Lyot stop, the segment gaps in the Keck primary mirror, which are less important, and mask imperfections, which are relatively minor (Vargas Catalan et al. 2016, in press). Nor does it address where the different leakages end up in the focal plane ($L_s$ yields a diffraction pattern, $L_p$ distorts the diffraction pattern, and $L_w$ produces speckles) or their interference effects, as neither affects the integrated leakage. As interference between scattered and diffracted light leads to enhanced semi-static "pinned" speckles along diffraction rings (Bloemhof et al. 2001), reducing the diffracted component with the vortex is important to reducing such bright speckles, and in providing an improved starting point for PSF-subtraction and speckle-suppression algorithms (Borde & Traub 2006, Lafreniere et al. 2007, Soummer et al. 2012, Amara & Quanz 2012, Martinache et al. 2014, Gomez Gonzalez et al. 2016).

The terms in Equation 1 are straightforward to estimate. The d/D ratio is not well defined for the hexagonal architecture of the Keck telescope, but for a 2.6 m central obscuration, a 10 m diameter primary has $L_s$ = 0.07, while the use of the inscribed telescope aperture diameter of 9.018 m would yield $L_s$ = 0.08. As to $L_w$, first generation AO systems typically provide wavefront rmses ($\lambda\varphi_{rms}/2\pi$) of ~ 200 to 250 nm, for which Eqn. 4 implies $S$ ~ 0.84 to 0.90 at L', giving $L_w$ ~ 0.1 to 0.16. Wavefront quality should thus limit the rejection somewhat more than the presence of a secondary mirror. Nevertheless, with $L_w$ and $L_s$ within a factor of two of each other, the stellar rejection enabled by a simple vortex coronagraph and the current Keck AO system are at compatible levels.

Pointing errors must then be kept small enough for the $L_p$ leakage to be smaller than the other two terms. Of these, $L_s$ is a constant for a given secondary mirror configuration, while $L_w$ follows the Strehl ratio, but can be reasonably constant during stable atmospheric conditions. Therefore, not only the level, but also the stability of the pointing leakage is very important, especially in reaching the smallest possible angles, and in enabling more effective use of PSF-subtraction algorithms in post-processing. As Strehl variations of 1% or so would be about the best that one could reasonably expect for a first generation AO system, we apply the same 1% leakage variation criterion to $L_p$, yielding a pointing accuracy requirement of $\frac{\sqrt{8}}{10\pi}\frac{\lambda}{D}$, or 7 mas at L' on a 10 m telescope.

What contrast can then be expected? Allowing for the fact that planet searches are conducted somewhat off-axis, where the ideal PSF is lower than the PSF peak by more than a factor of 100 beyond the first Airy ring (Born and Wolf 2011), and that the leakage estimates given above are PSF-integrated leakages (meaning that the peak flux in the typically ring-shaped post-vortex images will be further reduced due to being more spread out), and finally, that PSF differencing and calibration techniques should bring another one to two orders of magnitude of contrast improvement (e.g., Marois et al. 2008, Lafreniere et al 2009, Serabyn et al. 2010, Vigan et al. 2015, Gomez Gonzalez et al. 2016), best off-axis L' contrasts on the order of $10^{-5}$ should be feasible. However, as already mentioned, thermal background noise sets the ultimate attainable contrast level in the L' band. For example, in an hour-long observation with the Keck II telescope, the NIRC2 camera yields a 5 σ background noise level of L' ~ 18 mag, implying an ultimate 5 σ contrast level of ~ $10^{-4}$ for stars of L' ~ 8 mag in an hour. As the ultimate contrast in the background-limited regime improves linearly with stellar flux, an ultimate achievable contrast of ≈ $10^{-5}$ should be attainable for stars of L' ~ 5.5 mag. As usual, how close to the star this ultimate performance level is reached depends on the efficacy of the pointing correction and wavefront-error-reduction steps taken. Even so, as L'-band contrasts are equivalent to much lower young-jovian masses than at H-band, this is a promising performance regime.

## 3. Implementation of the NIRC2 vortex coronagraph

A coronagraph is more than simply a mask located in a focal plane (Lyot 1938, Sivaramakrishnan et al. 2001). Effective stellar rejection requires optimization of the full optical system, including the downstream Lyot stop, and the upstream adaptive optics (AO) system, which must provide a stable and high quality PSF centered on the focal-plane mask. This section describes the steps involved in implementing the Keck vortex coronagraph and in optimizing the combined Keck AO/NIRC2 system for L'-band coronagraphy. The overall development was of course greatly eased by the use of the existing AO system and camera.

The long-serving Keck AO system (Wizinowich et al. 2000) is based on a 349-actuator Xinetics deformable mirror (DM), and typically provides Strehl ratios of order 50-70% in the H/K bands for sufficiently bright natural guide stars, and significantly higher values (80-90%) at L'. However, as coronagraphy requires high-quality PSFs, one of the first steps necessary was to realign the Keck II AO system to eliminate a small amount of image elongation that had been present at L' band. This realignment was able to reduce the L' PSF ellipticity from 20% to less than 1% (Fig. 1).

For laboratory tests and system optimization steps, a new visible/infrared light source (Thorlabs SLS202) and a single-mode infrared optical fiber (Thorlabs ZrF$_4$) were used, both of which operate from the visible to L', thus easing co-registration of the visible-wavelength wavefront-sensor and L' science-camera beam trains. However, as the optical fiber is multi-mode below 2.3 μm, the system cannot currently be coaligned to the diffraction limit, and switching between short-wave and long-wave fibers remains necessary for wavefront optimization (e.g., for image sharpening, which is currently carried out in the NIR).

NIRC2 was the obvious camera of choice because it is already optically configured to operate as a coronagraph (with an internal focal plane containing a selection of opaque focal plane masks and a downstream pupil plane with a Lyot stop selection). We thus installed two IR subwavelength-grating vortex phase masks, referred to as annular groove phase masks (AGPM, Mawet et al. 2005), into NIRC2's internal focal plane. Two masks were installed for redundancy, to use different parts of the array, and to increase the possibility that one of them would allow operation in the M-band as well. The two masks are described in Vargas Catalan et al. (in press), and both produce output beams of topological charge 2, i.e., beams with phase wraps about the center of $2 \times 2\pi$ (Swartzlander 2009). In a laboratory coronagraphic configuration with a clear circular input pupil, both provided an attenuation for L'-band light of a factor larger than 100 (Fig. 2). (However, the rejection on the telescope will be lower due to the on-axis secondary; Sect. 2). Moreover, while both of the installed masks are optimized for L', some rejection is also expected in the longer-wavelength M band (Fig. 2). The two vortex phase masks were mounted in a new Al plate that took the place of the previous pinhole plate in the NIRC2 internal focal plane. The clear field-of-view transmitted by each mask, as mounted in the plate, is ~ 6" in the 10 mas-pixel NIRC2 camera (use of the finest NIRC2 plate scale is recommended for optimal pointing correction). The masks are installed such that the incident light first impinges on the side carrying a microstructured anti-reflective layer, with the vortex side being downstream. To install this vortex plate, the NIRC2 camera was taken off line, warmed, opened and cooled back down in March of 2015. No anomalies occurred during this process, but the NIRC2 distortion solution was substantially modified by the operation, requiring a recalibration (Service et al. 2016).

While the Lyot stop is also key to coronagraph optimization, the Lyot stops are much less accessible than the focal plane masks in the cryogenic NIRC2 camera, and so to minimize risk to this facility instrument, no Lyot stop masks were changed out. We thus instead simply rely on the best Lyot stop available, i.e., the one that yielded the lowest leakage in measurements obtained after the vortex mask installation. This turned out to be the inscribed circular Lyot stop, which is (partially) seen in Fig. 3, and which corresponds to a diameter on the primary of 8.72 m. Rotational co-alignment of this mask to the telescope pupil (Fig. 3) is achieved by means of the AO system's K mirror. Even so, as Fig. 3 indicates, some leakage remains near the telescope struts, especially near the contact points of the struts with the central blockage. The position of the telescope pupil relative to the Lyot stop is also quite sensitive to AO system and K-mirror alignment, and so is not stable either from run to run or with K-mirror rotation (i.e., some nutation is present). The pupil to Lyot stop co-alignment thus needs to be optimized prior to each observing run. Indeed, the Lyot stop mismatch and nutation are currently the primary performance limitations. For observations, the pupil is fixed with respect to telescope alt-az coordinates (pupil tracking mode). The combination of the vortex phase mask and the inscribed Lyot stop provided a rejection of our on-axis single-mode L' source (with no central obstruction in the beam) of a factor of several hundred, consistent with earlier laboratory performance tests of similar masks.

The response to off-axis point sources of the ideal Keck-II-telescope/NIRC2-vortex/Lyot-stop combination was simulated numerically. Fig. 4 shows the resultant "peak transmission," which is computed as the flux ratio integrated in a disk of diameter 1 $\lambda/D$ (thus excluding residual flux that has been moved out of the PSF core by passing close to the vortex center). Such calculations yield an effective IWA of 125 mas (Fig. 4), i.e., about 50% larger than the formal IWA of $\approx$ 80 mas that results for a clear circular aperture of a similar size in the case of integration over the full PSF.

When observing stars with this system, raw post-coronagraphic stellar focal-plane images show an uneven "string of pearls" of light at small radii (Fig. 1, center). In theory, round aperture boundaries yield uniform rings or doughnuts of residual light, while a hexagonal aperture yields a symmetric string of pearls (Fig. 1 right). The asymmetric lumpy ring seen when observing stars is thus presumably due to residual alignment errors. Reducing this leakage further would require a circular pupil mask at the DM, and also a correspondingly optimized Lyot stop.

Achieving stable pointing onto the vortex mask was one of our top priorities, as this is critical to accurate PSF subtraction, which enables reaching contrasts at or near the background limit in to the smallest angles. We therefore took several steps to optimize the pointing onto the vortex, from using the most relevant error signal, to the use of an optimized processing algorithm, to using the most appropriate actuation. The optimal pointing error signal is provided by the post-vortex stellar image on the science camera, as this image is directly impacted by pointing errors relative to the vortex mask itself, which leave a residual (asymmetric) doughnut of starlight on the science camera. For centrally blocked pupils, manual interpretation of the doughnut asymmetry is challenging, and can in practice be quite slow, so we automated the conversion of the post-coronagraphic PSF asymmetry on the NIRC2 science camera into an error signal to be sent to the AO system via the Quadrant Analysis of Coronagraphic Images for Tip-Tilt Sensing (QACITS) algorithm (Huby et al. 2015, 2016). Deduced corrections at the few mas level are then applied as centroid offsets to the Keck AO wavefront sensor, whence they are sent to the AO system's fast tip-tilt mirror in closed loop. With the finer NIRC2 pixel plate scale (9.971 mas/pixel; Service et al. 2015), the measured pointing stability of the QACITS loop is 3 mas rms, exceeding our nominal requirement. The primary goal of QACITS is to remove slow pointing drifts, and so the timescale for corrections is modest, roughly 30 sec.  In addition to drift removal, QACITS is also used for initially centering the star onto the vortex, in a much faster and more reproducible manner than manual alignment allows. Finally, the QACITS pointing loop serves to obviate the need for active differential angular refraction (DAR) correction between the science and AO tip-tilt sensing wavelengths while tracking, as keeping the star centered on the vortex is the actual goal. The DAR correction is thus used only for initial acquisition, and DAR tracking is turned off during the rest of an observation.

Wavefront error minimization is also required for deep starlight rejection. The Keck II AO system provides the bulk of this correction, including a standard Gerchberg-Saxton image-sharpening algorithm to reduce non-common path errors. Finally, low-order wavefront error terms such as focus errors onto the vortex mask plane (as opposed to onto the detector focal plane, which may not be exactly conjugate) and pupil shear errors on the Lyot mask, are typically examined visually in the pupil plane, in order to minimize obvious stellar leakage due to these errors (e.g., Fig. 3). Fig. 3 also shows slight leakage in the segment gaps, but at our relatively modest contrast levels, this is not an issue. Further reduction of scattered starlight is possible within the control region of the DM with speckle suppression techniques, and initial steps with speckle nulling are described in Bottom et al. (2016).

Finally, we note that the optimization of the coronagraphic system is not yet complete, and a number of further improvements are either in progress or under consideration. As such, the initial contrast performance is not particularly relevant, and so is not included here, as it would not represent the final capability of the new observing mode. This observing mode has also not yet been facilitized, as rapid development was the first priority. Thus further development steps remain, including testing the system at M band, converging on a stable and user-friendly facility observing mode, and developing and optimizing appropriate speckle suppression techniques.

**4. Initial Demonstration Results: HIP 79124 B**

As an initial demonstration of the capabilities of the vortex coronagraph, a number of young stars in the nearby Sco-Cen star-formation region (distance = 120 - 150 pc; de Zeeuw et al. 1999) were observed. Aperture masking interferometry had recently revealed close, faint companions to several Sco-Cen stars (Hinkley et al. 2015), providing very good test cases for the new system. With only single-epoch prior observations, the association of these faint neighbors with the brighter host stars remains to be confirmed, but the chance of random alignments at such close proximity was deemed rather low (Hinkley et al. 2015). In any case, physical association is irrelevant to the goal of demonstrating the ability to resolve faint objects from much brighter ones.

We observed both HIP 79124 and HIP 78233 on 2015 June 9 and again on 2016 April 13. Hinkley et al. (2015) lists the relevant stellar parameters for these stars, including spectral types of A0 and F0, and Wise W1 magnitudes of 6.96 and 7.64, respectively. Because of the improved vortex observing mode implemented by the time of the second observing run, the second data set was of much higher quality, and so only that data is discussed hereafter. The two stars are separated on the sky by ≈ 3.5°, and our observations alternated between the two stars so that we could use each of them to provide calibration PSFs for the other. We carried out two observation cycles, and switched between stars every 20 to 30 minutes. Because of significant readout overheads, total integration times were 11.5 min (23 images) and 18 min (36 images) on HIP79124 and HIP78233, respectively. Each image was the sum of 40 internally co-added frames, each of 0.75 sec integration time.

The data were reduced by means of reference-star differential imaging (RDI), using an algorithm based on principal component analysis (PCA), implemented in the Vortex Image Processing (VIP) package (Gomez Gonzalez et al., submitted). This algorithm uses all individual frames obtained on the reference star to build a low-rank approximation of the target star PSF using PCA. Angular differential imaging was not used, due to the slow parallactic rotation for these sources and the small radial offsets of interest. Background emission was subtracted from the target and reference frames using blank sky measurements taken just after the on-source integrations.

The final reduced image of HIP 79124 (Fig. 5) clearly shows an object significantly fainter than the primary at a separation of 186.5 ± 2 mas and position angle 246.9° ± 0.6°. The positional error bars were determined by injecting a series of fake companions around the star, and calculating the median of the retrieved positional error bar distribution (Gomez Gonzalez et al. 2016). The flux ratio, determined using short non-coronagraphic images, and corrected for a coronagraph transmission of 70% at the measured separation (Fig. 4) is $\Delta L' = 4.2 \pm 0.1$ mag. Our measured separation, position angle and contrast are close to those (177 ± 3 mas, 242° ± 1°, and 4.30 ± 0.10 mag, repectively) of Hinkley et al. (2015), thus confirming the aperture masking interferometry detection of this $\approx 100 - 200$ $M_{Jup}$ object. As far as we are aware, this is the first direct image of HIP79124 B. For a stellar distance of 123 pc (Hinkley et al. 2015), our observed offset corresponds to a projected separation of 23 AU. We note that this close-in detection required the small IWA of the vortex coronagraph, RDI, and the QACITS pointing stabilization loop, which together enabled PSF suppression at very small angles (~ 2.1 $\lambda/D'$, where D' is the diameter of the inscribed Lyot stop). In contrast, using the same observing strategy but without pointing stabilization in June 2015, only a marginal detection of the companion (at about the same position and flux) was possible.

In the slightly more than six years since the observations of Hinkley et al. (2015), even the small proper motion of HIP79124 [(-7.13 ± 0.79, -22.72 ± 0.60) mas/yr] yields a significant displacement, specifically by right ascension and declination offsets of -42.9 ± 4.8 mas and -136.8 ± 3.6 mas, respectively. If HIP79124 B were a background object, it would thus be displaced by this amount in the opposite direction, i.e., mainly toward the north by a very significant fraction of the original separation vector. This is definitely not seen, as our offset vector agrees with that of Hinkley et al. (2015) to an order of magnitude higher accuracy, implying that HIP79124 B is indeed physically associated with HIP79124 A.

However, our measured offset for HIP79124 B does differs by a *small* amount from that of Hinkley et al. (2015). The net displacement, accounting for the errors in both measurements, is 18 ± 4 mas, which is at the 4.5 σ level, suggesting orbital motion. At a distance of 123 pc, this angular displacement corresponds to a projected linear displacement of 2.2 AU. Two measurements are insufficient to determine an orbit, but the direction of the displacement suggests counterclockwise motion. A uniform circular orbit of a radius equal to the observed projected separation of 23 AU (which need not be the case), gives an orbital period of ~ 70 yr, while several degrees of angular displacement in 6 yr suggests a period on the order of 440 yr. Further observations will obviously be needed to confirm motion and to better constrain the orbital parameters.

## 5. Summary and Outlook

The new L' vortex coronagraph on the Keck II telescope provides a new small-angle high-contrast observational capability for the observatory. The performance of this mode is still being developed and characterized, but it can clearly already be used to detect companions very near the stellar position. The ultimate contrast performance obtained in long integrations will be addressed in the future (Bottom et al., in prep.), but because of the relative brightness of young Jovians at L', the sensitivity in terms of planet masses should be comparable to shorter wavelength ExAO high contrast systems. As both GPI and SPHERE are located in the southern hemisphere, the Keck vortex coronagraph provides a complementary small-angle, high-contrast system in the northern hemisphere, along with the LBT and SCExAO vortex coronagraphs (Defrere et al. 2014, Jovanovic et al. 2015).

While the main science case for this new system is very nearby young hot Jovian exoplanets, as demonstrated here, the small IWA capability of this system also enables direct observation of companions in the nearest star forming regions such as Sco-Cen and Taurus. This coronagraph thus may be able to provide data on the radial locations at which giant planets are actually being formed, of importance in discriminating between core-accretion (planet formation generally closer in) and disk instability (planet formation generally farther out) scenarios, as well as the hot- and cold-start planet formation models (e.g., Baraffe, Chabrier & Barman 2010). The system should also be quite capable in other observational areas, e.g., in observing emission from circumstellar disks and planets forming therein, where the L'-band has already shown potential (e.g., Kraus & Ireland 2012). Here again the small IWA is crucial, both in accessing more numerous sources out to large distances, and in accessing angles in to the ice line region for nearer sources. Finally, this system should also enable sensitive searches for exoplanets around nearby M stars, where again small angles are crucial (Mawet et al. 2016). Thus, this new observational mode should enable progress in a significant number of key topics in the formation of planetary systems.

Finally, we note that further development steps in the direction of high contrast capabilities are of course possible. Although the current Keck AO system and a simple vortex coronagraph configuration are at compatible performance levels, both could conceivably be upgraded, the AO system to higher order correction, and the coronagraph to a configuration better suited to a centrally obscured aperture (Mawet et al. 2013, Serabyn, Liewer & Mawet 2016), or to a segmented aperture telescope (Ruane et al. 2016). Such performance improvements could also extend high-contrast operation to shorter wavelengths, and so to yet smaller angles. This development is thus potentially only a first step toward an expanded suite of high contrast capabilities at the observatory.

**Figure Captions**

**Fig. 1:** Measured L'-band pre- (left) and post-vortex-coronagraph (center) focal plane PSFs, both normalized by the maximal value of the non coronagraphic PSF. Points to note are the round central lobe on the left, the six-pointed ring of residual light in the center, and the asymmetry of that ring, the latter likely due to residual misalignments. The rejection is seen directly by comparing these two images. Right: model calculation of the expected rejection, for perfect alignment.

**Fig. 2:** Broadband peak rejection ratios measured in the lab for the two AGPMs installed in NIRC2 (see Vargas Catalan et al. [2016] for further details). The lengths of the solid lines represent the passbands of the filters used. The dashed curves illustrate the theoretical performance of these two AGPMs based on our best estimation of their grating parameters (grating depth d, line width w, and side-wall angle α).

**Fig. 3:** Two images of the NIRC2 Lyot stop plane (on a linear intensity scale) showing the residual pupil-plane leakage. In the left hand image, the K-mirror has rotated the pupil image away from the secondary supports. In the right hand image, the brighter regions at the bases of the secondary supports can be seen. One can also see faint outlines of the hexagonal telescope panels, but at these modest rejection levels the panel gaps are not a concern.

**Fig. 4:** Model calculation of the azimuthally-averaged off-axis peak transmission of the Keck L' vortex coronagraph. The transmission is estimated here as the ratio of the flux integrated inside a disk of diameter 1 λ/D, which takes into account the fact that the shape of the image is affected by the coronagraph, in particular at very small angular separations. Due to the secondary mirror, the hexagonal aperture shape, and the loss of light out of the PSF core, the half-power peak transmission occurs at 125 mas.

**Fig. 5:** Final reduced image of HIP79124 B obtained with the Keck L-band vortex coronagraph. The Hinkley et al. (2015) location for HIP79124 B is indicated by an "X".

**Acknowledgements**

We thank the Keck Science Steering Committee for their approval of this project, and especially J. Cohen for her support in getting this project off the drawing board. We also thank the director of the Caltech Optical Observatories, Shri Kulkarni, for providing test time for this new mode. Finally, we thank the W.M. Keck Observatory staff for their able and enthusiastic assistance with the observations. The data presented herein were obtained at the W.M. Keck Observatory, which is operated as a scientific partnership among the California Institute of Technology, the University of California and the National Aeronautics and Space Administration. The Observatory was made possible by the generous financial support of the W.M. Keck Foundation. The research leading to these results has received funding from the European Research Council under the European Union's Seventh Framework Programme (ERC Grant Agreement n. 337569), the French Community of Belgium through an ARC grant for Concerted Research Action, and the Swedish Research Council (VR) through project grant 621-2014-5959. Part of this work was carried out at the Jet Propulsion Laboratory, California Institute of Technology, under contract with NASA.

# References

Absil, O., Milli, J., Mawet, D., Lagrange, A.-M., Girard, J., Chauvin, G., Boccaletti, A., Delacroix, C., & Surdej, J. 2013, A&A 559, L12

Amara, A. & Quanz, S.P. 2012, MNRAS 427, 948

Baraffe, I., Chabrier, G., & Barman, T. 2010, Rep. Prog. Phys. 73, 016901

Baraffe, I., Chabrier, G., Barman, T. S., Allard, F., & Hauschildt, P. H. 2003, A&A 402, 701

Biller et al. 2014, ApJL 792, L22

Beuzit, J-L et al. 2008, in Proc SPIE 7014, 701418

Bloemhof, E.E., Dekany, R.G., Troy, M., & Oppenheimer, B.R. 2001, ApJL 558, L71

Borde, P.J. and Traub, W.A. 2006, ApJ 638, 488

Born, M. & Wolf, E. 2011, Principles of Optics, 7$^{th}$ ed., Cambridge

Bottom, M. et al., in Proc. SPIE 9909, 990955

Burrows, A., Marley, M., Hubbard, W. B., Lunine, J. I., Guillot, T., Saumon, D., Freedman, R., Sudarsky, D., & Sharp, C. 1997, ApJ 491, 856

Defrère, D. et al. 2014, Proc. SPIE 9148, 91483X

Delacroix, C., Absil, O., Forsberg, P., et al. 2013 A&A 553, A98

de Zeeuw, P. T., Hoogerwerf, R., de Bruijne, J. H. J., Brown, A. G. A., & Blaauw, A. 1997, AJ 117, 354

Foo, G., Palacios, D.M., & Swartzlander, G.A. Jr. 2005, Opt. Lett. **30**, 3308

Fortney, J.J., Marley, M.S., Saumon, D., & Lodders, K. 2008, ApJ 683, 1104

Forsberg, P. & Karlsson, M. 2013, Diamond and Related Materials 34, 19

Gomez Gonzalez, C. A., Absil, O., Absil, P.-A., Van Droogenbroeck, M., Mawet, D., & Surdej, J. 2016, A&A 589, A54

Hinkley, S. Kraus, A.L., Ireland, M.J. et al. 2015, ApJL 806, L9

Huby, E., Baudoz, P., Mawet, D., & Absil, O. 2015, A&A 584, A74

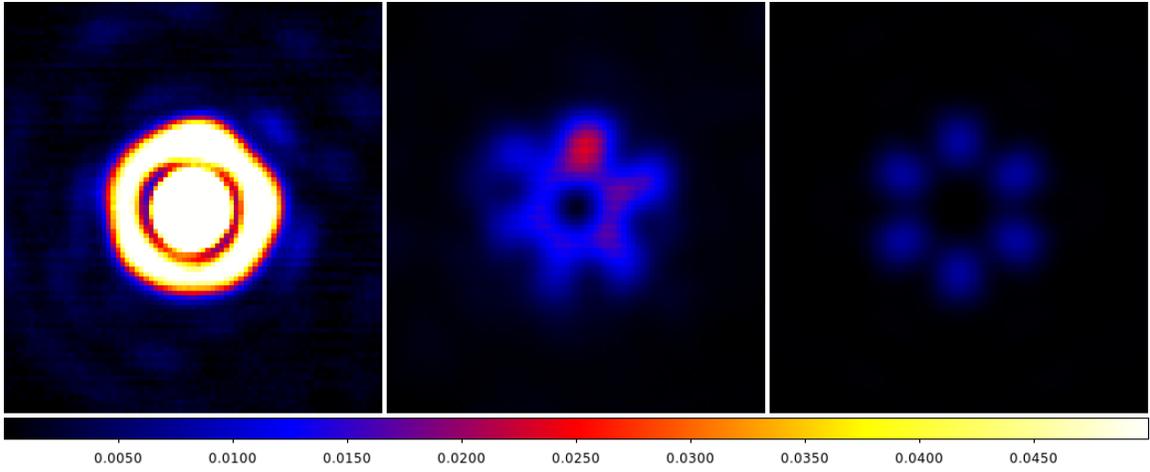

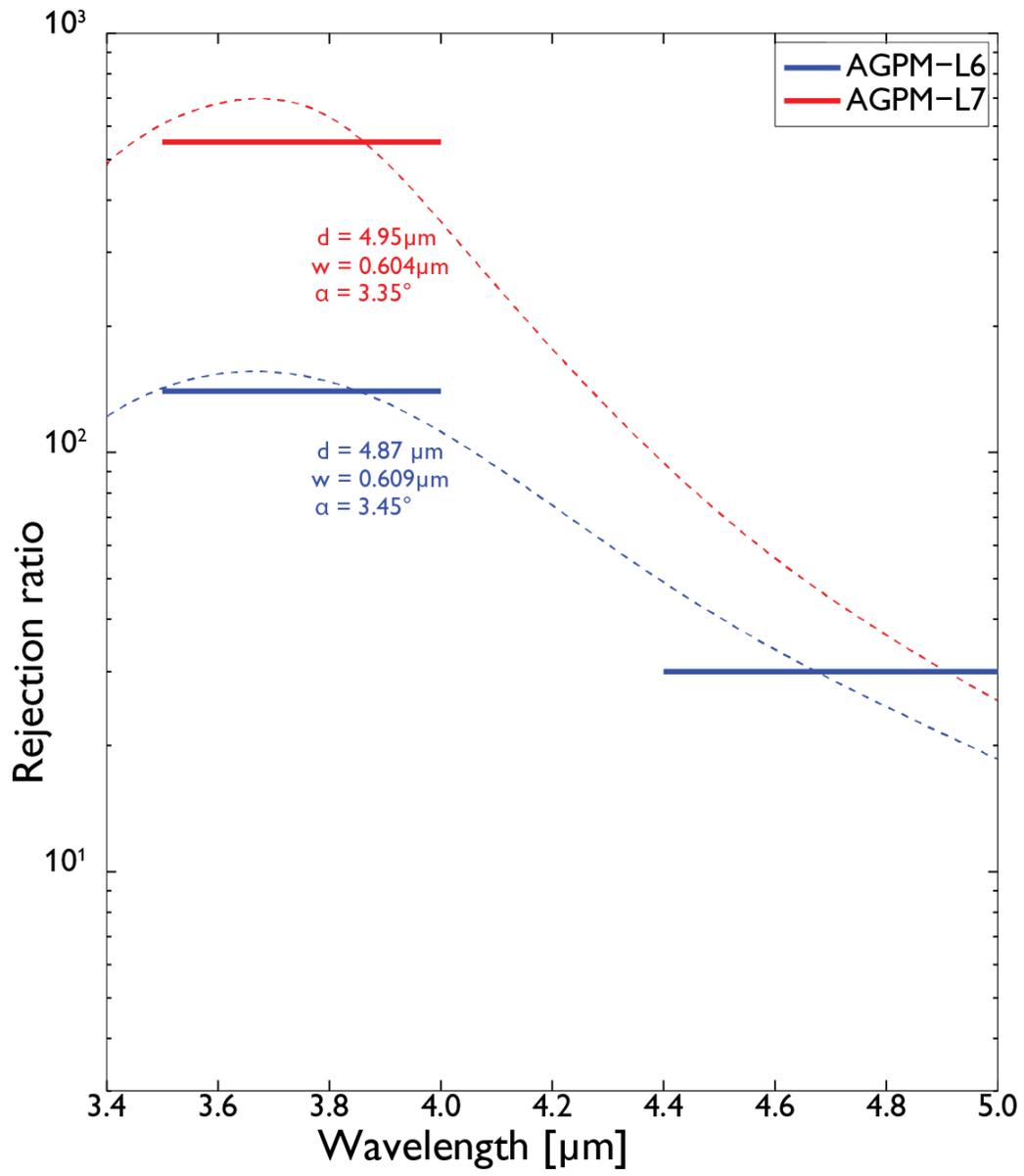

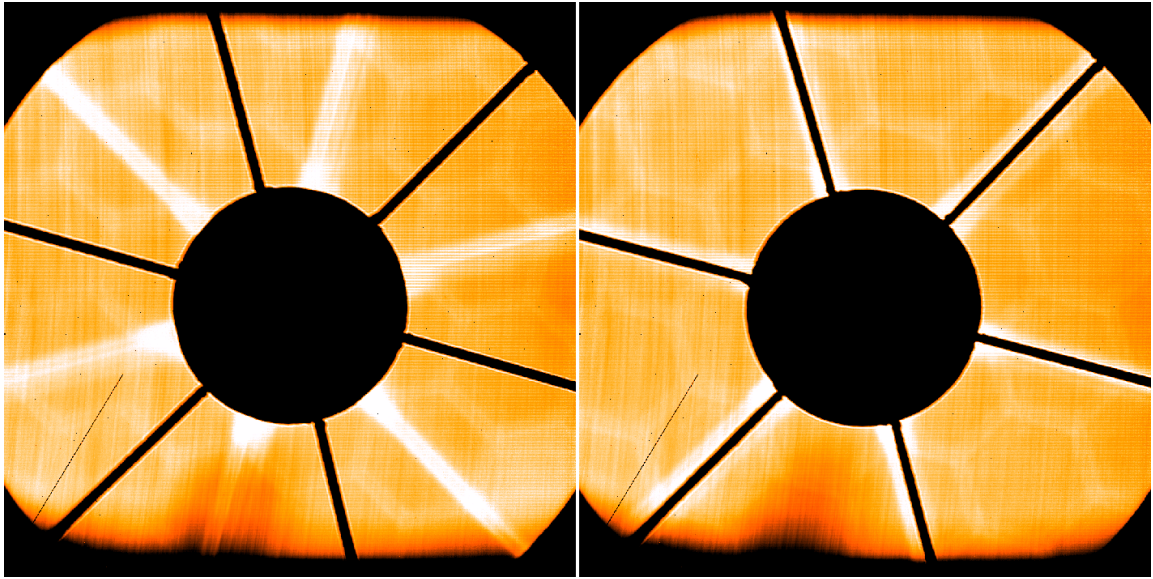

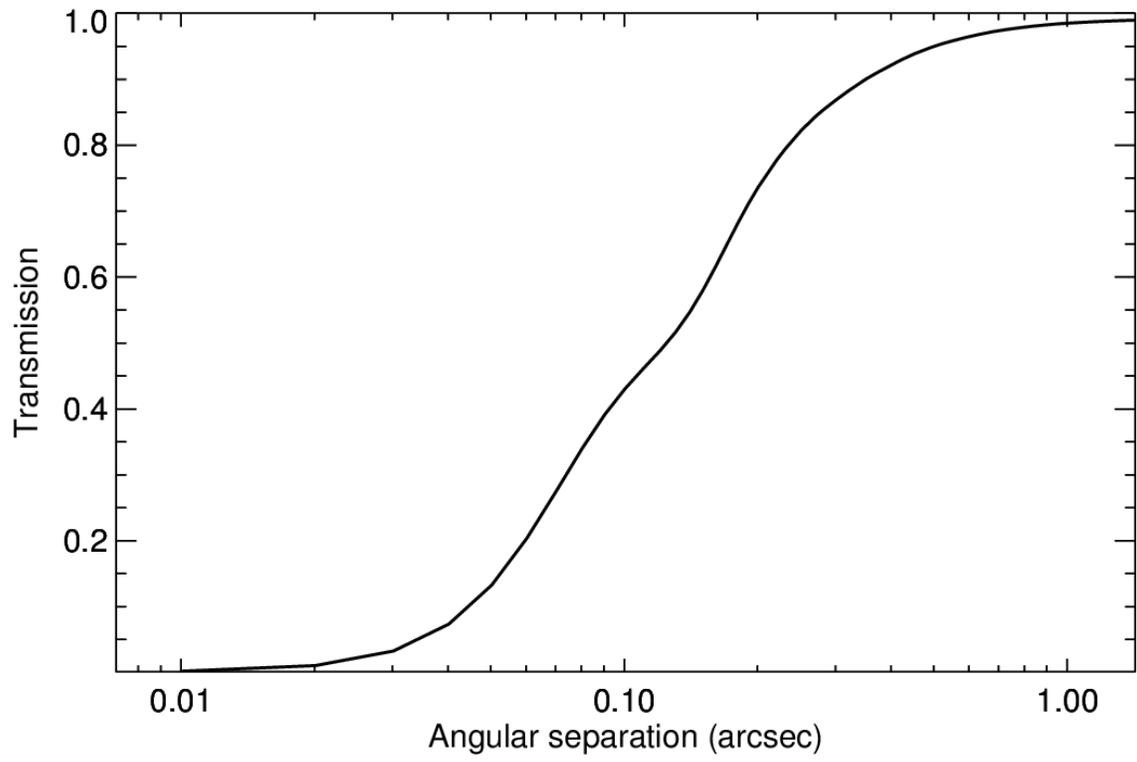

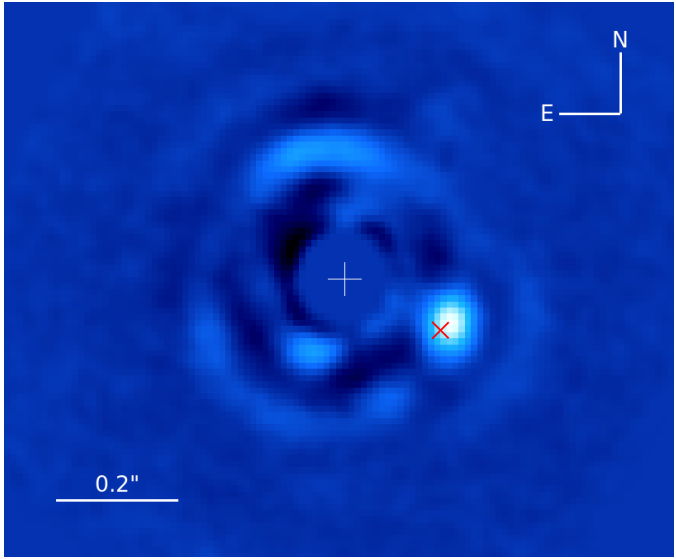